\documentclass[superscriptaddress,aps,showpacs,nofootinbib,preprintnumbers,twocolumn]{revtex4}
\usepackage{amsmath,amssymb}
\usepackage{graphicx}
\usepackage{dcolumn}
\usepackage{bm}

\def \gsim{\mathrel{\vcenter
     {\hbox{$>$}\nointerlineskip\hbox{$\sim$}}}}

\newcommand{\beq}{\begin{equation}}
\newcommand{\eeq}{\end{equation}}
\newcommand{\beqa}{\begin{eqnarray}}
\newcommand{\eeqa}{\end{eqnarray}}
\newcommand{\beqar}{\begin{eqnarray*}}
\newcommand{\eeqar}{\end{eqnarray*}}

\begin{document}

\preprint{UG-FT-184/05, CAFPE-54/05, ROMA-1403-05}

\title{
TeV gravity at neutrino telescopes
}

\author{J. I. Illana}
\email{jillana@ugr.es}
\author{M. Masip}
\email{masip@ugr.es}
\affiliation{
CAFPE and
Depto.~de F{\'\i}sica Te\'orica y del Cosmos, Universidad de Granada,
18071 Granada, Spain}

\author{D. Meloni}
\email{meloni@roma1.infn.it}
\affiliation{
INFN and 
Dipto.~di Fisica, Universit\`a degli Studi di Roma "La Sapienza", 
00185 Rome, Italy}


\begin{abstract}

Cosmogenic neutrinos reach the Earth with energies 
around $10^{9}$ GeV, and their interactions with matter 
will be measured in upcoming experiments (Auger, IceCube). 
Models with extra dimensions and the fundamental scale at 
the TeV could imply signals in these experiments.
In particular, the production of microscopic black holes 
by cosmogenic neutrinos 
has been extensively studied in the literature. Here we make a complete 
analysis of gravity-mediated interactions at larger distances, 
where they can be calculated in the eikonal approximation. 
In these processes a neutrino of energy $E_\nu$ 
interacts elastically with a parton
inside a nucleon, loses a small fraction $y$ of its energy, and 
starts a hadronic shower of energy $y E_\nu \ll E_\nu$. We analyze 
the ultraviolet dependence and the relevance of graviton emission
in these processes, and show that they are negligible. 
We also study the energy distribution of cosmogenic events 
in AMANDA and IceCube and the possibility of multiple-bang events.
For any neutrino flux, 
the observation of an enhanced rate of neutral current 
events above 100 TeV in neutrino telescopes could
be explained by TeV-gravity interactions. 
The values of the fundamental scale of gravity  
that IceCube could reach are comparable to those to be 
explored at the LHC.

\end{abstract}

\maketitle

\section{Introduction}

Cosmogenic neutrinos appear in any scenario proposed
to explain the most energetic cosmic rays. In particular, 
if the observed air showers  
of up to $10^{11}$~GeV \cite{Anchordoqui:2002hs} are 
produced by primary protons, in their way to the Earth 
these protons will interact with the cosmic microwave background 
(CMB) photons and produce pions:
\beq
p + \gamma_{2.7{\rm K}} \to \Delta^+\to n+ \pi^+\; (p+\pi^0)\;.
\label{eq1}
\eeq
The flux of cosmogenic neutrinos would then be created in the 
decay of the charged pions, and it will appear correlated with 
observable fluxes of nucleons and photons (see \cite{semikoz} for a 
recent review).

Cosmogenic neutrinos are of great interest as probes of new TeV 
physics because they provide very large 
center of mass energies. In addition, the relative effect of new 
physics on the weakly interacting neutrinos is larger than on quarks 
or charged leptons, making it easier to see deviations. At these  
energies the new physics may be able to {\it compete} with the 
weak interactions and provide signatures that could be detected in 
deeply penetrating air showers and neutrino telescopes.

In particular, one expects that at transplanckian energies 
gravity dominates over all the other interactions. 
This will be the case when a cosmogenic neutrino interacts with
a terrestrial nucleon in models with extra dimensions and the 
fundamental scale $M_D$ at the TeV \cite{ADD}.
The possibility of black hole (BH) formation \cite{BH0} by 
cosmogenic neutrinos has been discussed by several groups 
\cite{feng,ringwald,halzen,olinto,casadio,Giddings:2004xy,emparan}. 
These analyses are based on
a {\it geometric} cross section, which assumes 
gravitational collapse if the neutrino interacts at
impact parameter distances smaller than the Schwarzschild radius $R_S$
of the system. The collapse involves strongly coupled
gravity and is not calculable perturbatively, but if $R_S\gg M_D^{-1}$
($\sqrt{s}\gg M_D$) one expects
that the estimate will not be off by any large 
factors \cite{Giddings:2004xy}. It is found,
however, that the $\nu N$ cross section is dominated by the 
low-$x$ region, with $\sqrt{s}$ at the parton level 
close to $M_D$, and most of the BHs produced will have a radius
$R_S\sim M_D^{-1}$. 
In this regime the amount
of gravitational radiation emitted during the collapse or the topology
of the singularity are important effects that add uncertainty to
the geometric estimate.

Here we study the gravitational interaction
at larger distances, where it can be calculated
using the eikonal approximation \cite{emparan,us,eik,riccardo}.
This approximation involves linearized
gravity and is not affected by the uncertainties in the cross section
for BH formation. In the next Section we show 
that, for the typical energy $E_\nu$ of cosmogenic neutrinos,
the eikonalized $\nu N$ cross section 
depends very mildly on how the theory is completed in the
ultraviolet (UV), at energies
around $M_D$. We also show that the amount of gravitational 
radiation emitted during the scattering is small.
In these processes the neutrino interacts at 
impact parameters larger than $R_S$ and therefore with a larger
cross section than for BH production. At the parton level
the neutrino scatters elastically and transfers a small fraction 
$y$ of its energy to a quark or a gluon, which starts then 
a hadronic shower of energy $yE_\nu$. Three main
features characterize these processes and
distinguish them from standard model or BH events.
First, the shower has a typical energy much smaller than the 
energy ($10^{8}$ to $10^{11}$ GeV) of the incoming neutrino. 
Second, a charged lepton is {\it never} produced in the starting
point of the shower. 
Finally, the neutrino is {\it not} destroyed in the interaction,
it keeps going with essentially the same energy and may interact
again. We show in Section III that neutrino telescopes are 
then ideal experiments to observe these elastic
processes: they are designed to detect hadronic
showers of energy down to 100 TeV $\ll 10^{9}$ GeV (below
100 TeV the atmospheric background dominates), and 
their big volume
(1 km$^3$ in IceCube \cite{icecube}) would favor multiple-bang events.
In the final section we discuss and summarize our results.
Our analysis here completes our work in \cite{us}, where aspects like
the UV dependence of the eikonal amplitude, gravitational 
bremsstrahlung,
or multiple-bang events in neutrino telescopes were not discussed.

\section{TeV gravity}

The simplest picture of TeV gravity includes only two free parameters:
the value of the higher-dimensional Planck scale $M_D$, and the number
$n$ of compact dimensions
where gravity propagates. A third parameter, the
(common) length $2\pi R$ of the $n$ dimensions, could be deduced from the
4-dimensional Newton constant $G_N\equiv M_P^{-2}$:
\beq
G_D= (2\pi R)^n G_N = {(2\pi)^{n-1}\over 4 M_D^{n+2}}\;.
\label{eq2}
\eeq

At processes below $M_D$ the model-independent signature of
extra dimensions is graviton emission.
The amount of energy radiated
would be proportional to the accessible phase space or, in the
Kaluza-Klein (KK) picture, to the number of KK modes of mass
below the center of mass energy. In this type of experiments
for a given $n$ one sets bounds on $R$ and then deduces the
limits on $M_D$.
From collider experiments one obtains $M_D\ge 1.4\;(1.0)$ TeV
for $n=2\;(\ge 3)$ \cite{colliders}, whereas from
SN1987A the bounds go up to 22 TeV for $n=2$ \cite{SN}.
One should keep in mind, however, that the gravitons emitted in
the supernova explosion have a KK mass below $\approx 50$ MeV. The simple
picture with two extra flat dimensions 
could be modified above 50 MeV, for example, with
four more dimensions at $R'\sim (100$~GeV)$^{-1}$, which would bring
the fundamental scale of gravity down to 1 TeV without affecting
the physics in the supernova. It could also be that some other 
mechanism (a warp factor in \cite{Giudice:2004mg}) gives an extra mass
of order $\ge 50$ MeV to the KK excitations, invalidating all the
bounds based on supernovas.

The bounds obtained from transplanckian collisions are
complementary in the sense that given $n$ they are a direct probe of
$M_D$, and $R$ is then adjusted in order to reproduce $G_N$.
At energies above $M_D$ and impact parameters smaller than $R$
the collision is a pure higher-dimensional process independent
of the compactification details that fix the value of
the effective Newton constant. The transplanckian
collision does not {\it see} that the extra dimensions
are compact, they could be taken infinite
with no effect on the cross section.

\subsection{Neutrino-parton amplitude}

The TeV gravity model should be embedded in a string theory, which
would relate $M_D$ with the string scale $M_S$. In the simplest
set-up \cite{peskin} the standard
model (SM) fields (open strings) would be attached to
a 4-dimensional brane, whereas gravity (closed strings) would propagate
in the whole $D$-dimensional space. In this case
\beq
M_D^{n+2}={8 \pi\over g^4} M_S^{n+2}\;,
\label{eq3}
\eeq
with $g$ the string coupling. The transplanckian regime
corresponds then to energies above the string scale, where
any tree-level amplitude becomes very {\it soft}. In the
ultraviolet string amplitudes go to zero exponentially
at fixed angle and, basically, only the forward (long-distance)
contribution of the graviton survives
(the forward contribution of the SM gauge bosons also
survives, but it is subleading above $M_D$ due to the smaller
spin of the vector bosons). This is precisely the regime
where the eikonal approximation is valid.

Let us consider the elastic collision of a neutrino and a
parton that exchange $D$-dimensional gravitons (see
\cite{emparan,riccardo} for details).
The eikonal amplitude ${\cal A}_{\rm eik}(s,t)$ resums the infinite
set of ladder and cross-ladder diagrams.
It is reliable as far as
the momentum carried by the gravitons is smaller than
the center of mass energy or, in terms of the fraction
of energy $y=(E_\nu-E'_\nu)/E_\nu$ lost by the incoming neutrino,
if $y=-t/s\ll 1$ ($s$ and $t$ refer to the Mandelstam
parameters at the parton level). In this limit the
amplitude is independent of the spin of the colliding particles.
Essentially, ${\cal A}_{\rm eik}$ is the exponentiation of the Born
amplitude in impact parameter space:
\beq
{\cal A}_{\rm eik}(s,t)={2 s\over i}
\int {\rm d}^2b\; e^{i\mathbf{q}\cdot\mathbf{b}}\;
\left(e^{i\chi (s,b)}-1\right)\;,
\label{eq4}
\eeq
where $\chi (s,b)$ is the eikonal phase
and $\mathbf{b}$ spans the (bidimensional) impact parameter space.
The Born amplitude corresponds to ${\cal A}_{\rm eik}(s,t)$ in 
the limit of small $\chi (s,b)$ and, therefore, the eikonal
phase can be deduced from the Fourier transform to
impact parameter space of ${\cal A}_{\rm Born}(s,t)$:
\beq
\chi(s,b) = \frac{1}{2s}\int\frac{d^2q}{(2\pi)^2}
{\rm e}^{-i{\bf q}\cdot{\bf b}}{\cal A}_{\rm Born}(s,q^2)\ .
\label{eikphase}
\eeq
Our Born amplitude comes from the $t$-channel exchange
of a higher-dimensional graviton:
\beq
{\cal A}_{\rm Born}=-{s^2 \over M_D^{n+2}}
\int {{\rm d}^n\;q_T\over t-q_T^2} \;,
\label{eq5}
\eeq
where the integral over momentum $q_T$ along the extra dimensions
(equivalent to the sum over KK modes) gives an UV
divergence if $n\ge 2$. 
The {\it magic} of the eikonal amplitude is
that it will be well defined despite we obtain it from an UV
dependent Born amplitude.
To understand that, let us first evaluate $\chi (s,b)$ using
dimensional regularization. The Born amplitude becomes
\beq
{\cal A}_{\rm Born}(s,t)=\frac{s^2}{M^{n+2}_D}
\pi^{\frac{n}{2}}(-t)^{\frac{n}{2}-1}\Gamma\left(1-\frac{n}{2}\right)\ ,
\eeq
which implies
\beqa
\chi(s,b)&=&
\frac{\pi^{\frac{n}{2}-1}\Gamma\left(1-\frac{n}{2}\right) s}
{4M^{n+2}_D}\int^\infty_0 dq\ q^{n-1} J_0(qb) \nonumber \\
&=&\frac{1}{b^n}
\frac{(4\pi)^{\frac{n}{2}-1}}{2}\Gamma\left(\frac{n}{2}\right)
\frac{s}{M^{n+2}_D}
\equiv \left(\frac{b_c}{b}\right)^n.\quad
\eeqa
Although the eikonal phase diverges at $b=0$, the 
amplitude in Eq.~(\ref{eq4}) is insensitive to that: the contributions 
from the region 
$b\ll b_c$ are quickly oscillating and tend 
to cancel. 
${\cal A}_{\rm eik}(s,q)$ can be written
\beq
{\cal A}_{\rm eik}(s,q)=4\pi s b_c^2\; F_n(b_c q)\;,
\label{eik2}
\eeq
with
\beq
F_n(u)=-i
\int_0^\infty {\rm d}v\;v\; J_0(uv)
\left( e^{iv^{-n}} -1 \right)\;,
\label{eq7}
\eeq
where $q=\sqrt{-t}$, and the integration variable is $v=b/b_c$
(see Fig.{\ref{fig1}).
For $q<b_c^{-1}$ this integral is dominated by impact parameters
around $b_c$, and for $q>b_c^{-1}$ by
a saddle point at $b_s=b_c (n / qb_c )^{(1/n+1)}$.

\begin{figure}
\includegraphics[width=\linewidth]{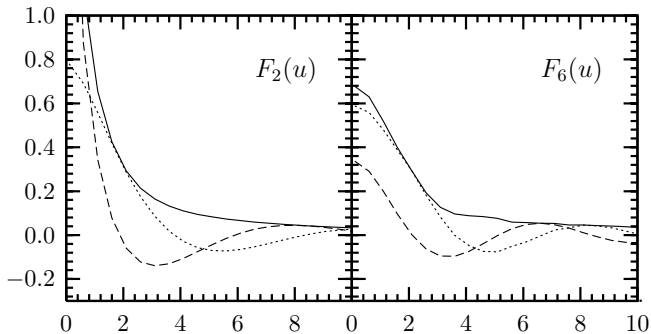}
\caption{
The real (dashed) and imaginary (dotted) parts and the moduli (solid)
of the functions $F_n$  in the eikonal amplitude (\ref{eik2}) for 
$n=2,\ 6$. \label{fig1}}
\end{figure}

\subsection{\label{subsect2.2}Cutoff dependence of the amplitude}

Let us now use an UV cutoff $\Lambda$ to regularize the Born
amplitude in Eq.~(\ref{eq5}). This may be more {\it physical} than dimensional
regularization since $D$-dimensional gravity must be completed
(embedded in a consistent theory) at energies around $M_D$.
For example, in the simple brane-world setting mentioned
above all the KK excitations of the graviton with mass 
(or $q_T^2$) larger than the string scale decouple 
exponentially \cite{Antoniadis:1993jp}, an effect that would 
mimic a cutoff. In any case, we can use the cutoff to estimate 
the UV dependence of these eikonalized processes.

The Born amplitude becomes
\beqa
{\cal A}_{\rm Born}(s,q) &=&
\frac{s^2}{M_D^{n+2}}\frac{2\,\pi^{n/2}}
{\Gamma\left(\frac{n}{2}\right)}
\,\int_0^\Lambda
dq_T\, \frac{q_T^{n-1}}{q_T^2+q^2}  \nonumber \\
&=&\frac{s^2}{M_D^{n+2}}\frac{2\,\pi^{n/2}}
{\Gamma(n/2)}
\, q^{n-2} I_n(\Lambda/q)\;,
\eeqa
where $I_n(\Lambda/q)$ diverges like $(\Lambda/q)^{n-2}$ with 
the cutoff (the divergence is logarithmic
for $n=2$). It is now straightforward to find
\beqa
\chi (s,b)=\frac{\pi^{\frac{n}{2}-1} s}{2M_D^{n+2}
\Gamma (n/2)}\frac{1}{b^n}\;\int_0^{\Lambda b}
d \xi\ \xi^{n-1}\,K_0(\xi)
\label{coff}
\eeqa
where $K_0(\xi)$ is a modified Bessel function of the second kind.
Expanding (\ref{coff}) in powers of $\Lambda b$ we obtain
\beqa
\chi (s,b)=\left(\frac{b_c}{b}\right)^n
\left[1 - \sqrt{\frac{\pi}{2\Lambda b}}
\, {\rm e}^{-\Lambda\,b}\, A_n(\Lambda b) \right]\ ,
\eeqa
with $A_n(\Lambda b)$ approaching the constant 
$2^{2-n}/\Gamma^2(n/2)$ for large values of 
$\Lambda b$. This expression tells us that the cutoff
introduces corrections to the eikonal phase which are relevant
only at impact parameters $b\le \Lambda^{-1}$. This region
in $b$ gives a negligible contribution to ${\cal A}_{\rm eik}(s,q)$
(see Section \ref{example}).

\subsection{Non-linear corrections and soft graviton emission}

The eikonal amplitude in Eq.~(\ref{eik2}) is well defined 
for all values of $s$ and $q$. However, 
as $q$ (or $y=q^2/s$) grows nonlinear corrections
(H diagrams) become important \cite{riccardo}. The relevance
of H diagrams implies a regime with 
strong gravitational coupling and important 
graviton emission (soft bremsstrahlung).
The strong coupling can be expected just by inspecting 
the eikonal amplitude, since for $-t/s\approx 1$
the saddle point $b_s$ that dominates the integral in 
impact parameter space 
approaches the
Schwarzschild radius $R_S$ \cite{emparan} of the system:
\beq
R_S=\left[{2^n\pi^{n-3\over 2}\Gamma\left({n+3\over 2}\right)
\over n+2}\right]^{1\over n+1}
\left({s\over M_D^{2n+4}}\right)^{1\over 2(n+1)}\;.
\label{eq8}
\eeq
A process with typical impact parameter $b\leq R_S$  
will not be properly described by the eikonal amplitude,
since nonlinear corrections will be of order one. 
On the other hand, 
in eikonal processes with $y\ll 1$ the main contribution
to the amplitude in Eq.~(\ref{eq4}) comes from impact 
parameters much larger than $R_S$, 
where nonlinear effects are small (see Section \ref{example}).

Soft graviton emission is also a consequence of nonlinear
couplings, it appears as an imaginary contribution to the 
eikonal phase corrected by H diagrams ($\chi_H$) \cite{Amati:1990xe}.
This contribution is of absorptive type, it damps the elastic
cross section showing that a Bloch-Nordsieck mechanism is at
work. For a given value of $b$, the
average number $N_{\rm soft}$ of gravitons radiated during the 
scattering can be read directly from $\chi_H$ 
\cite{Amati:1990xe,riccardo}:
\beq
N_{\rm soft}={\rm Im}\;(\chi_H)\approx 
\left( {b_r\over b}\right)^{3n+2}\;,
\eeq
where
\beq
b_r\equiv \left( b_c^n R_S^{2n+2}\right)^{1\over {3n+2}}\approx
\left( G_D^3 s^2\right)^{1\over {3n+2}}\;.
\eeq
Therefore, the typical (transverse) momentum radiated will be
$Q\approx N_{\rm soft} b^{-1}$. Notice that to obtain the energy lost
by the incoming neutrino this momentum must be boosted 
to the nucleon rest frame.
In an eikonal scattering the dominant impact parameter distance 
is $\langle b\rangle\approx b_s$. Both 
$Q\approx b_s^{-1}\approx M_D (yM_D^2/s)^{1/2n+2}$
and the number of gravitons 
$N_{\rm soft}\approx y^{(3n+2)/(2n+2)}(s/M_D^2)^{(n+2)/(2n+2)}$
decrease for decreasing values of $y$,
implying that for $y\ll 1$ the amount of 
gravitational radiation during the scattering is small
(see below a numerical example).

On the other hand, 
in a collision at $\langle b\rangle\approx R_S$ 
one expects a large fraction of energy transferred from the
neutrino to the parton, a large scattering angle,
and a significant fraction of energy lost to radiation.
At these and smaller values of $b$ one would also expect 
black hole (BH) formation 
\cite{BH0,feng,ringwald,halzen,olinto,casadio,Giddings:2004xy,emparan}. 
It has been shown, however, that a number of factors
(angular momentum, charge,
geometry of the trapped surface, radiation before the collapse)
make a precise estimate difficult, specially for 
light BHs of mass just above $M_D$.

\subsection{\label{example}Numerical analysis of the $\nu N$ eikonal cross 
section}

To understand the relative relevance of the different scales
and processes involved, in this Section we will consider the 
scattering of a $10^{10}$ GeV neutrino with a nucleon 
$N=(n+p)/2$. The $\nu N$
center of mass energy is in this case 
$\sqrt{s}=\sqrt{2m_N E_\nu}=141$ TeV. We will take 
$n=2$ or $n=6$ extra dimensions and a fundamental scale $M_D=1$ TeV.
The transplanckian regime will then
include the partonic processes of
energy $\sqrt{\hat s}=\sqrt{x s} > M_D$, {\it i.e.,} 
$x > 5\times 10^{-5}$. To evaluate the cross sections we will 
use the {\tt CTEQ5} parton distribution 
functions (PDFs) \cite{Lai:1999wy}, 
which are available both for {\sf fortran} and {\sf Mathematica} codes. 
We will base our analysis on the kinematical variable 
$y=(E_\nu-E'_\nu)/E_\nu$, which fixes $q^2=y\hat s$ and
the dominant impact parameter distance $\langle b\rangle$ 
in the eikonal process ($\langle b\rangle \approx b_s$ if 
$q>b_c^{-1}$ or $\langle b\rangle \approx b_c$ if 
$q<b_c^{-1}$).
We will evaluate the PDFs at this dominant distance 
($\mu=\langle b\rangle^{-1} $).

We find (see Fig.~\ref{fig2}) that the differential 
cross section 
\beqa
\frac{d\sigma^{\nu N}_{\rm eik}}{dy}=\int^1_{M^2_D/s} dx\ xs\ 
\pi b^4_c \left|F_n(b_cq)\right|^2
\sum_{i=q,\bar{q},g}f_i(x,\mu)\;
\eeqa
grows as $y$ decreases \cite{emparan}. For example, 
for $n=2\;(6)$ it is a
factor of 265 (62) larger at $y=10^{-3}$ than at
$y=0.1$. 
The small $y$ region corresponds to long distance processes
where the neutrino interacts with a parton and transfers only a small
fraction of its energy.
This region is less important for $n=6$ than for $n=2$ 
extra dimensions, since then gravity {\it dilutes} faster and becomes
weaker at long distances.
On the other hand, values of $y$ close to 1 mean shorter distance
interactions. Using Eq.~(\ref{eik2}) 
we can evaluate the contribution to 
the cross section from different regions in impact parameter
space. We obtain, for example, that for a $\nu N$ process
with $y=0.5$ a $52\%$ of
the eikonal cross section comes from impact parameters
$b<R_S$ if $n=2$ (or a $71\%
$ if $n=6$).
In these processes with $y\approx 1$ the eikonal amplitude will be 
corrected by nonlinear contributions of the same order.
Therefore, we will  
use the eikonal amplitude to evaluate
elastic processes with  
$y<y_{\rm max}\approx 0.2$ only. Our results will depend very
mildly on the actual value of $y_{\rm max}$, since the 
bulk of the cross section comes from the small $y$
region. For example, for $n=2$ the eikonal $\nu N$ cross section 
for processes with $10^{-6}\le y\le y_{\rm max}$ is 
$\sigma^{\nu N}_{\rm eik}=1.97 \times 10^{-2}$ mb if $y_{\rm max}=0.4$ 
or $\sigma^{\nu N}_{\rm eik}= 1.91 \times 10^{-2}$ mb if $y_{\rm max}=0.1$. 
For $n=6$ the values of the cross would change with $y_{\rm max}$
from $7.5 \times 10^{-3}$ mb to $6.2 \times 10^{-3}$ mb. 

\begin{figure}
\includegraphics[height=0.7\linewidth]{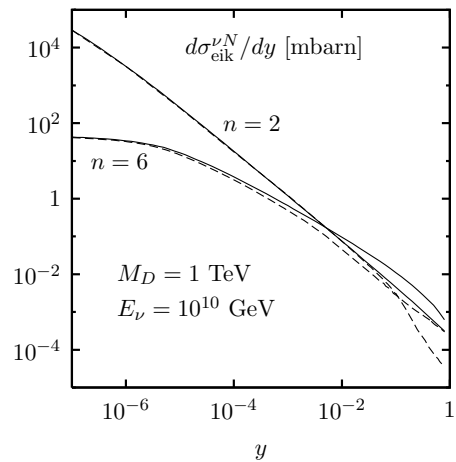}
\caption{Differential cross section 
${\rm d}\sigma^{\nu N}_{\rm eik}/{\rm d} y$ 
of a $10^{10}$~GeV neutrino 
for $n=2,\ 6$ and an UV cutoff $\Lambda\rightarrow \infty$
(solid) and $\Lambda=M_D$ (dashed).
\label{fig2}}
\end{figure}

The numerical relevance of the UV cutoff 
(see Section \ref{subsect2.2})
is expressed in Fig.~\ref{fig2}, where we plot 
${\rm d}\sigma^{\nu N}_{\rm eik}/{\rm d} y$ for $\Lambda\rightarrow \infty$
(solid lines) and for $\Lambda=M_D$ (dashed line). In the later case KK modes 
of the graviton heavier than $M_D$ are decoupled. We find that 
the $\nu N$ differential cross section changes less than a 10\% for 
$10^{-5}<y<10^{-2}$, so the cutoff dependence of the eikonal
amplitude in these processes is not important.

Let us now consider the geometric cross section 
(at the parton level) 
$\sigma_{\rm BH}=\pi R_S^2$, with $R_S$ given in Eq.~(\ref{eq8}).
$\sigma_{\rm BH}$ includes all
the processes (elastic and inelastic) 
at impact parameter distances $b$ below $R_S$, and it 
can be used to estimate the rate of black hole formation.
As explained before, 
eikonal scatterings of small $y$ will be dominated
by values of $b$ larger than $R_S$. Therefore, the 
overlapping between these soft eikonal 
processes and the processes in the (inclusive)
geometrical cross section will be negligible. It
is then justified to consider two types of transplanckian
($\hat s>M_D^2$) processes:
elastic (long-distance) processes where the
neutrino transfers to the partons a small fraction 
$y<y_{\rm max}$ of its
energy and keeps going, and shorter distance ($b<R_S$) 
{\it hard} processes
where the neutrino loses in the collision most of its energy,
possibly collapsing into a BH. 
To estimate the relative frequency 
of these two processes when a $10^{10}$ GeV neutrino scatters
off a nucleon, 
we can compare the eikonal cross section
$\sigma^{\nu N}_{\rm eik}$ with $y$ integrated 
between $10^{-5}$ and $0.2$ 
with $\sigma^{\nu N}_{\rm BH}$. For $n=2\;(6)$ we obtain
$\sigma^{\nu N}_{\rm eik}= 1.94 \times 10^{-2}$ mb ($6.88 \times 10^{-2}$ mb) 
and $\sigma^{\nu N}_{\rm BH}= 9.82 \times 10^{-4}$ mb 
($4.07 \times 10^{-3}$ mb), {\it i.e.}, the
neutrino will have 12.5 (1.64) interactions in which it transfers to the
nucleon between 100 TeV  and $2\times 10^{9}$ GeV of energy, per
each short distance (black hole) interaction. The total 
energy lost by the neutrino in these 12.5 (1.64) interactions is
\beqa
E_{\rm eik} &=&\frac{1}{\sigma^{\nu N}_{\rm BH}}
\displaystyle 
\int^{y_{\rm max}}_{{100\;{\rm TeV}\over E_\nu}} dy\; y E_\nu\;
\frac{d\sigma^{\nu N}_{\rm eik}}{dy} \nonumber \\
&=& 1.00 \times 10^{9} \;{\rm GeV}\;\;
(5.04 \times 10^{8} \;{\rm GeV).}
\eeqa

Finally, let us comment on the amount of gravitational 
energy radiated 
in these eikonalized scatterings. In a parton
process of energy $\sqrt{x s}$ and inelasticity $y$ the 
energy lost to radiation in the c.o.m. 
frame is 
\beq
E^*_{\rm rad}(x,y)\approx \min\left\{
{1\over \langle b\rangle} 
\left({b_r\over \langle b\rangle}\right)^{3n+2},
\sqrt{x s} \right\}\;,
\eeq
which in the nucleon at rest frame is
\beq
E_{\rm rad}(x,y)= E^*_{\rm rad} \sqrt{E_\nu\over 2xm_N}\;.
\label{erad}
\eeq
The average energy lost through soft bremsstrahlung
per each short distance gravitational interaction 
is then 
\beqa
\langle E_{\rm rad}\rangle =\frac{1}{\sigma^{\nu N}_{\rm BH}}
\int^{1}_{M_D^2/s} dx \int^{y_{\rm max}}_0 dy\;E_{\rm rad}
\;\frac{d^2\sigma^{\nu N}_{\rm eik}}{dx\;dy}
\eeqa
We find that, for $n=2\; (6)$, during the 12.5 (1.64) eikonal 
processes the $10^{10}$ GeV neutrino radiates soft gravitons with a 
total energy of $1.56 \times 10^{9}$ GeV ($4.90 \times 10^{8}$ GeV).

\section{Signals at neutrino telescopes}

The flux of cosmogenic neutrinos depends on the production
rate of primary nucleons of energy around and above the GZK
cutoff $E_{\rm GZK}$.
It will always appear correlated with proton
and photon fluxes that should be consistent, respectively,
with the number of ultrahigh energy events at AGASA and
HiRes \cite{Anchordoqui:2002hs} and with the
diffuse $\gamma$-ray background measured by EGRET
\cite{Sreekumar:1997un}.

We will base our analysis on the two neutrino fluxes 
described in \cite{semikoz} (solid and dashed lines in Fig.~\ref{fig3}).
The first one saturates the observations by EGRET, whereas for
the second one the correlated flux of $\gamma$-rays contribute
only a 20\% to the data, with the nucleon flux normalized
in both cases to AGASA/HiRes. The {\it higher}
flux predicts 820 downward neutrinos 
of each flavor with energy between
$10^8$~GeV and $10^{11}$~GeV per year and km$^2$, versus
370 for the {\it lower} one. We will also comment on the {\it minimal}
flux described in \cite{Fodor:2003ph} (dots in Fig.~\ref{fig3}), 
where the cosmic ray events above the GZK cutoff are 
assumed not to be protons \cite{domokos}. The proton 
events around and below $E_{\rm GZK}$ imply then  
just 35 downward neutrinos of each flavor with energy between
$10^8$~GeV and $10^{11}$~GeV per year and km$^2$.

\begin{figure}
\includegraphics[width=\linewidth]{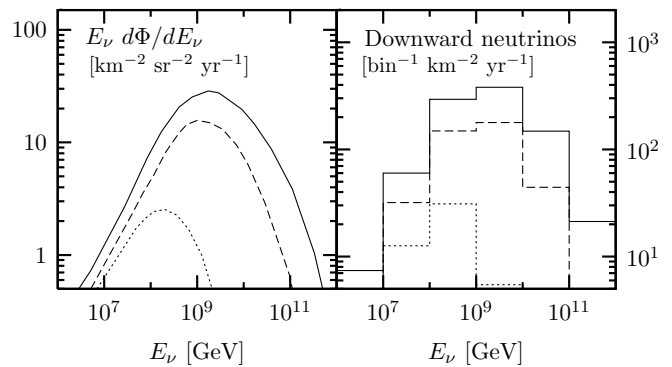}
\caption{
Left: cosmogenic neutrino fluxes referred in the text as
{\it higher} (solid), {\it lower} (dashed) and
{\it minimal} (dotted). We plot the fluxes for one flavor 
$\Phi=\phi_{\nu_\ell}+\phi_{\bar\nu_\ell}$ and assume that all
flavors have the same frequency. Right: 
corresponding number of downward $(0\le\theta_z<\pi/2)$
cosmogenic neutrinos of each flavor.
\label{fig3}}
\end{figure}

To observe a hadronic event inside the telescope, a 
cosmogenic neutrino must first {\it survive} as it crosses 
the atmosphere and the ice (or water) above the detector. 
Its typical interaction length in a medium of density $\rho$ is
\beq
L_0=\frac{1}{\rho N_A\sigma^{\nu N}}\ ,
\eeq
where $\sigma^{\nu N}=\sigma^{\nu N}_{\rm SM}+\sigma^{\nu N}_{\rm BH}$ is 
the total cross section to have an interaction that 
{\it destroys} the neutrino. It is usual to express the
length in terms of its depth: $x_0=\rho L_0$ ({\it i.e.} one
meter of water has a depth of 100 g/cm$^2$).
Notice that we include in
$\sigma^{\nu N}$ both the SM and the short distance 
gravitational interactions, but we ignore the
soft (eikonalized) gravitational interactions because
they take from the neutrino just a small fraction of
its energy (the distortion in the flux of neutrinos that reach the 
detector that these interactions produce is negligible).

A neutrino from a zenith angle $\theta_z$ must cross a column 
density of material
\beq
x(\theta_z)=\int_{\theta_z} dl\ \rho(l,\theta_z)\;.
\eeq
In practice, the path in the atmosphere is negligible and 
$x(\theta_z)$ is just the depth of the water or ice 
above the detector. The probability that it does not
interact before reaching the detector is then
\beq
P_{\rm surv}(E_\nu,\theta_z)={\rm e}^{-x/x_0}\ .
\eeq
Once in the detector, the probability of an event is
\beq
P_{\rm int}(E_\nu)\approx 1-{\rm e}^{-L\rho N_A\sigma^{\nu N}_{\rm int}}\ ,
\label{pint}
\eeq
where $L$ is the linear dimension of the detector and
$\sigma^{\nu N}_{\rm int}$ the total cross section. Therefore, the 
total number of events in the telescope in an observation
time $T$ is
\beqa
N=2\pi AT \int dE_\nu \sum_{\nu_i,\bar\nu_i}
\frac{d\phi_{\nu_i}}{dE_{\nu}}
\int d\cos\theta_z P_{\rm surv} P_{\rm int}\ ,
\eeqa
where $A$ is the detector's cross sectional area and $\phi_{\nu_i}$ 
the neutrino flux. Before a complete numerical analysis we
would like to discuss the possibility of multiple-bang 
events inside the detector.

\subsection{Multiple-bang events}

If $L$ is similar or larger than the interaction length $L_0$, 
then the neutrino may interact more than once inside the
detector. This is possible because the neutrino is not 
{\it destroyed} in the first eikonal interaction, it
keeps going with basically the same energy and can interact
again.

Let us assume that 
$\sigma^{\nu N}_{\rm int}\approx \sigma^{\nu N}_{\rm eik}\gg 
\sigma^{\nu N}_{\rm BH},\sigma^{\nu N}_{\rm SM}$ 
and let us neglect the amount of energy lost by the neutrino
in each interaction. It is straightforward to find the 
probability of having {\it exactly} $N$ interactions 
({\it bangs}) in a length $L$:
\beq
P_{N}(L)={\rm e}^{-{L/L_0}}\, \frac{(L/L_0)^N}{N!}\ .
\eeq
For example, the probability of having only one interaction 
would be $P_{1}(L)=\exp{(-L/L_0)} (L/L_0)$; for $L\ll L_0$ we
have $P_{1}(L)\approx L/L_0$, but for $L\gg L_0$ this amplitude
goes to zero (it is very unlikely to have only one
interaction). Given $L$, the most probable number of 
interactions is
$N=L/L_0$, which is also the average number of interactions:
\beq
\langle N\rangle =\sum_{N=1}^{\infty}\,N\,P_{N} = {L\over L_0}\ .
\eeq

Notice that the probability of having any type of event, {\it i.e.},
at least one interaction, is (see Eq.~(\ref{pint})) 
\beq
P(L)=\sum_{N=1}^{\infty}\,P_{N}= 1-{\rm e}^{-L/L_0}\ ,
\eeq
whereas the probability that this event includes more
than one interaction (a multiple-bang event) would
be $P(L)-P_1(L)$:
\beq
P_{\rm mult}(L)= 1-{\rm e}^{-{L/L_0}}(1+{L/L_0})\ .
\eeq

Double-bang events could also be produced by SM or BH interactions.
Within the SM, the second bang would correspond to the decay of
the tau created in the first interaction. The probability that
this happens would be 
the probability that a $\nu_\tau$ has a SM charged current
interaction times the probability that the tau lepton decays
inside the detector. 
If the $\nu N$ interaction results into
a BH, its evaporation would also produce taus that could decay
inside the detector. For the double-bang tau event 
to be contained inside a detector like IceCube (1 km of length with  
125 m between strings), the energy of the tau lepton must be between
$2.5\times 10^{6}$ GeV and $10^{7}$ GeV \cite{halzen}.

\subsection{Numerical example}

Let us consider again a single neutrino of $E_\nu=10^{10}$ GeV.
Within the SM, its interaction length in ice is $L^{\rm SM}_0=440$ km. This
means that typically it could reach the center of AMANDA or IceCube 
(1.8 km below the antarctic ice \cite{icecube}) from angles 
$\cos\theta_z \ge -0.03$. If there are  $n=2\;(6)$ extra dimensions 
and $M_D=1$ TeV, the interaction 
length before a {\it hard} gravitational interaction would be 
just $L_0=17$ km (4 km), which corresponds to 
$\cos\theta_z \ge 0.11 \,(0.44)$.
For $M_D=2.8\;(4.5)$ TeV 
$\sigma^{\nu N}_{\rm BH}\approx \sigma^{\nu N}_{\rm SM}$
and $L_0\approx L^{\rm SM}_0$.

If the neutrino reaches the detector, within the SM the probability
that in $L=1$ km (the longitudinal dimension of IceCube) 
it starts a hadronic shower is $P^{\rm SM}_{\rm int}=2.2\times 10^{-3}$.
For a $\nu_\tau$ neutrino, 
the probability of an event with a tau lepton 
in the initial point of the hadronic shower is $1.6 \times 10^{-3}$.

If $M_D=1$ TeV
the probability of a short distance gravitational
event would be $P^{\rm BH}_{\rm int}=0.06$ if $n=2$ (or 0.22 for $n=6$). 
To find the probability
of a {\it soft} eikonalized interaction we need to evaluate
\beq
\sigma^{\nu N}_{\rm eik} = 
\int^{y_{\rm max}}_{y_{\rm min}} dy
\frac{d\sigma^{\nu N}_{\rm eik}}{dy}\;,
\eeq
with $y_{\rm max}=0.2$ 
and $y_{\rm min}=(100\;{\rm TeV})/E_\nu$ (the energy
of the shower should be above 100 TeV to avoid the atmospheric
background). The probability of an event
in a length $L$ would then correspond to 
$\sigma_{\rm int}\approx \sigma^{\nu N}_{\rm eik}$ in Eq.~(\ref{pint}),
which for $L=1$ km and $n=2\;(6)$ gives $P^{\rm eik}_{\rm int}=0.56\;(0.33)$.
This probability includes events with one bang:
$P^{\rm eik}_{1}=0.36\;(0.27)$, with two bangs:
$P^{\rm eik}_{2}=0.15\;(0.06)$, and with more than two bangs:
$P^{\rm eik}_{>2}=0.05\;(0.008)$.

As explained above, 
double-bang events could also be produced by SM interactions.
The probability that a $10^{10}$ GeV $\nu_\tau$ produces a 
tau lepton of energy between $2.5\times 10^{6}$ and $10^{7}$ GeV
is just $P^{\rm SM}_{2}\approx 6.8 \times 10^{-5}$.

\begin{figure*}
\includegraphics[width=0.3\linewidth]{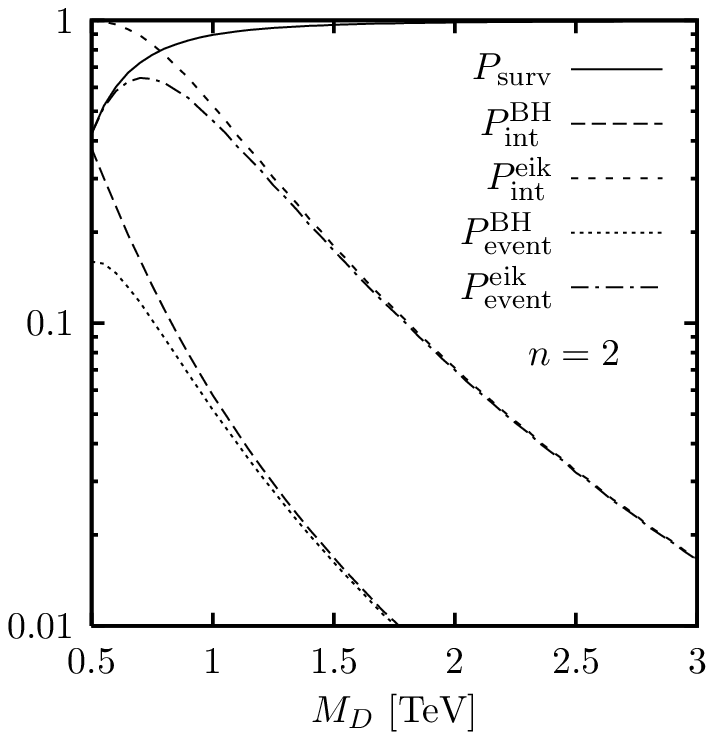}
\quad
\includegraphics[width=0.3\linewidth]{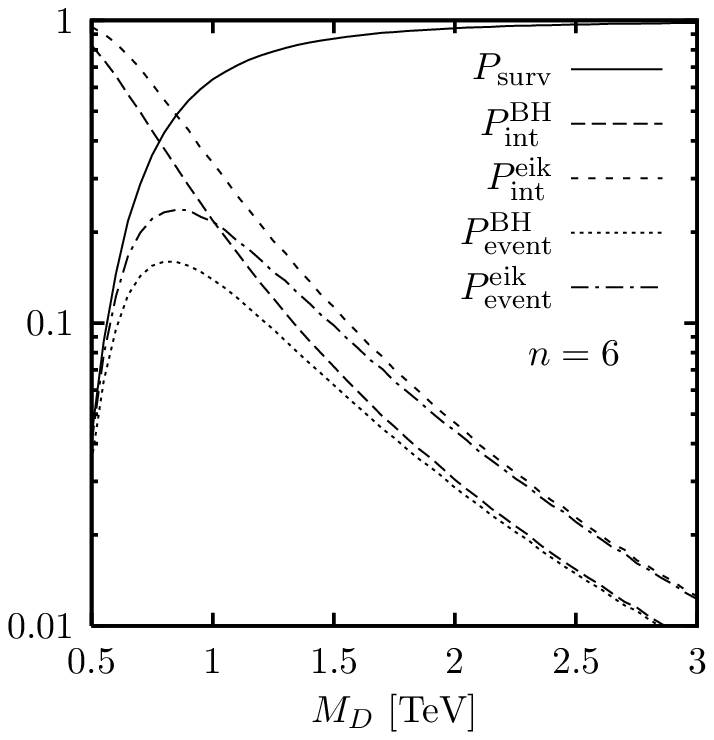}
\quad
\includegraphics[width=0.3\linewidth]{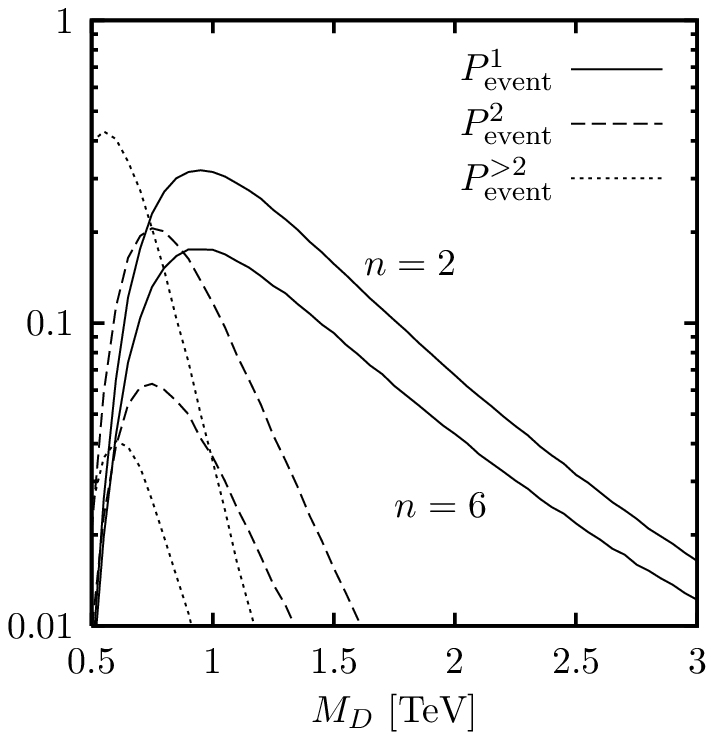}
\caption{
The different probabilities defined in the text for a $10^{10}$~GeV 
neutrino reaching IceCube from $\theta_z=0$ as a function of $M_D$
for $n=2$ and $n=6$. 
\label{fig4}}
\end{figure*}

In Fig.~\ref{fig4} we express the probability  $P_{\rm surv}$ 
that the $10^{10}$ GeV
neutrino survives through 1.8 km of ice to reach vertically 
IceCube for different values of
$M_D$ (for large values of $M_D$ $P_{\rm surv}\approx 
P^{\rm SM}_{\rm surv}\approx 1$). 
We also plot the probability that
if it has reached the detector it experiences 
a {\it hard} interaction
($P^{\rm BH}_{\rm int}$) or a {\it soft} observable interaction
($P^{\rm eik}_{\rm int}$) within a distance of $L=1$ km. The product
$P_{\rm event}=P_{\rm surv}P_{\rm int}$ would give the probability that 
the neutrino gives a signal in IceCube. We obtain that
this probability is larger for eikonal than for BH
events. We also find that it is maximal for $M_D\approx 0.8$ TeV.
For lower values of $M_D$ the neutrino tends to interact 
before reaching the detector, and for larger values it tends
to go through the detector without interactions. In Fig.~\ref{fig4}
we also plot the probability of an eikonal event that includes 
only one bang ($P^{1}_{\rm event}$), two bangs ($P^{2}_{\rm event}$), 
or more than two bangs ($P^{>2}_{\rm event}$). If $M_D\gsim 1.5$ TeV
the probability of more than one bang would be very small.

Within a distance $L$ the average energy lost 
by the neutrino in eikonal 
interactions and radiated through
gravitational bremsstrahlung would be, respectively, 
\beqa
\langle E_{\rm eik}\rangle=L\rho N_A 
\int^{y_{\rm max}}_0 dy\;y E_{\nu}
\;\frac{d\sigma^{\nu N}_{\rm eik}}{dy}\;,
\eeqa

\beqa
\langle E_{\rm rad}\rangle=L\rho N_A 
\int^{1}_{M_D^2/s} dx \int^{y_{\rm max}}_0 dy\;E_{\rm rad}
\;\frac{d^2\sigma^{\nu N}_{\rm eik}}{dx\;dy}, 
\eeqa
with  $E_{\rm rad}$ given in Eq.~(\ref{erad}).
For $L=1$ km and $n=2\;(6)$ the $10^{10}$ GeV neutrino will lose 
$\langle E_{\rm eik}\rangle=6\times 10^7\;(1.2\times 10^8)$ GeV 
to hadrons and
$\langle E_{\rm rad}\rangle=9.2\times 10^7\;(1.2\times 10^8)$ GeV to 
gravitational radiation. This means
that, as it propagates in the detector, the energy loss in these soft 
processes is negligible. In a typical interaction length $L_0$
of a hard interaction (where the neutrino will lose most or
all of its energy) we find that 
$\langle E_{\rm eik}\rangle/E_\nu=0.10\;(0.05)$ and 
$\langle E_{\rm rad}\rangle/E_\nu=0.16\;(0.05)$.

Finally, let us study what is the typical energy of the
hadronic shower started by the $10^{10}$ GeV neutrino. 
We will increase $M_D$ to 2 TeV (to avoid double-bang events) and 
take $n=2\;(6)$. If
the neutrino reaches the detector at IceCube, the 
probability that it starts a shower of energy between
100 TeV and $0.2E_\nu$ is
$P_{\rm int}\approx 1-{\rm e}^{-L \rho N_A\sigma^{\nu N}_{\rm int}}$,
where $\sigma^{\nu N}_{\rm int}\approx \sigma^{\nu N}_{\rm eik}$ 
with $y_{\rm min}={100\;{\rm TeV}/E_\nu}$ and $y_{\rm max}=0.2$.
We obtain $P_{\rm int}=0.070\;(0.045)$. 
Now, the probability $\Delta P_{\rm int}$ 
that this event has an energy between
$E_1$ and $E_2$ is
\beq
\Delta P_{\rm int}=\frac{P_{\rm int}}{\sigma^{\nu N}_{\rm eik}}
\int^{E_2\over E_\nu}_{E_1\over E_\nu} dy
\frac{d\sigma^{\nu N}_{\rm eik}}{dy}\;.
\eeq
In Fig.~\ref{fig5} we divide the interval from 100 TeV to $2\times10^{9}$ GeV
in 5 bins and express the probability that the shower energy is
in each of the bins. We observe that the typical value is
between $10^5$ and $10^8$ GeV for $n=2$ and between $10^6$ and 
$10^9$ GeV for $n=6$.

\begin{figure}
\includegraphics[height=0.7\linewidth]{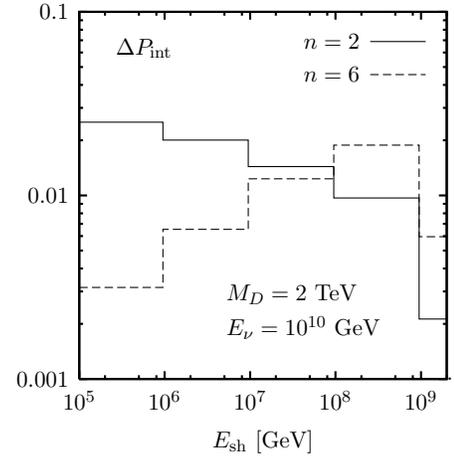}
\caption{
Probability that a $10^{10}$~GeV neutrino that reaches IceCube starts 
an eikonal shower in each interval of energies for $M_D=2$ TeV 
and $n=2,\ 6$. For example, the neutrino has a probability of 0.024
of starting a shower of energy between 100 and 1000 TeV if $n=2$.
\label{fig5}}
\end{figure}

\subsection{Bounds from air showers}

The fact that the typical energy $E_{\rm sh}=yE_\nu$ of the 
shower started in these processes is much smaller than 
the energy $E_\nu\approx 10^{10}$ GeV of the incoming 
neutrino has implications in air shower experiments. 
In particular, the absence of deeply penetrating showers 
in AGASA and Fly's Eye could exclude $\nu N$ cross sections 
between 0.01 and 1 mb. Notice, however, that these experiments 
are only sensitive to showers of very large energy, with 
$E_{\rm sh}$ around or above $10^9$ GeV. Since the eikonal 
cross sections that we are considering reach a large size 
only for low values of $y$, the typical showers that 
they produce would be {\it invisible} in AGASA 
and Fly's Eye. This is in contrast to processes 
like BH formation \cite{feng} or other processes 
of strongly interacting neutrinos \cite{domokos}, where most
of the energy of the initial neutrino would be transferred 
to the shower. 

A precise analysis of the
bounds on $M_D$ from eikonal processes in deep air shower
experiments can be found in \cite{us}.
The limits obtained there, between 1 TeV for $n=2$ and 
1.5 TeV for $n=6$, are essentially the same as the ones from
BH production in \cite{feng}.

\subsection{Cosmogenic neutrinos at AMANDA and IceCube}

Let us now study the total number of events at AMANDA (0.03 km$^2$
and a length of 700 m) and IceCube (1 km$^3$) 
for the neutrino fluxes in Fig.~\ref{fig3}. 

In the SM,
for the {\it higher} flux (910 downward cosmogenic neutrinos of each flavor
per year and km$^2$), 
we would expect 1.32 contained events per year in IceCube. 
Of those, 0.38 would come from a neutral current and
0.94 from a charged current (one third of the events of each 
lepton flavor).  
The distribution of energy of these events is given in
Fig.~\ref{fig6} 
(we show it despite the low statistics just for a comparison purpose).
Around 0.008 of the 0.31 tau events would decay 
and give {\it double bangs} inside the detector.
For the {\it lower} flux (around 410 cosmogenic neutrinos of 
each flavor per year
and km$^2$), we would expect just 0.50 SM events per year inside the
detector, 0.12 of them containing a tau lepton, and just 0.003
double-bang events.
The numbers for AMANDA can be easily obtained just multiplying
by a volume factor $V_{AM}/V_{IC}\approx 0.02$, namely 0.03 SM events 
per year for the {\it higher} flux and 0.01 for the {\it lower} flux. 

\begin{figure}
\includegraphics[height=0.95\linewidth]{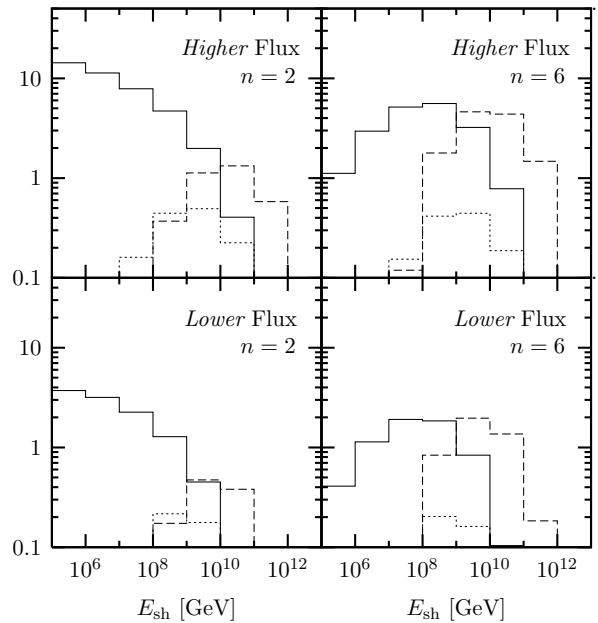}
\caption{
Energy distribution (events per bin) of the eikonal (solid), BH (dashed) 
and SM (dotted) events in IceCube per year for the {\it higher} and 
the {\it lower} 
cosmogenic fluxes, $M_D=2$~TeV and $n=2,\ 6$.
\label{fig6}}
\end{figure}

\begin{figure}
\includegraphics[height=0.95\linewidth]{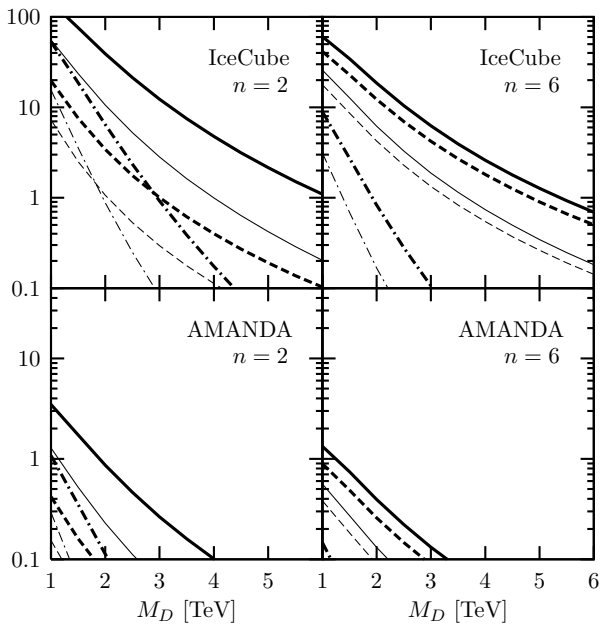}
\caption{
Contained events per year in IceCube and AMANDA for the
{\it higher} (thick) and the {\it lower} (thin) cosmogenic 
fluxes and $n=2,\ 6$.
We show eikonal (solid), multi-bang (dashed-dotted) and BH (dashed) 
events.
\label{fig7}}
\end{figure}

In a scenario with $n=2\;(6)$ extra dimensions, 
for the {\it higher} flux
we obtain a signal above the SM background 
(1.32 contained events per year in IceCube)
if $M_D\le 5.6\;(4.9)$ TeV, whereas for the {\it lower}
flux we have a signal above the 0.50 SM events 
if $M_D\le 4.8\;(4.5)$ TeV.

The event rate at IceCube and AMANDA for different
values of $M_D$ and a minimum energy of the shower of 100 TeV
is plotted in Fig.~\ref{fig7}.
We give the number of short distance (BH) and 
of soft (eikonal) events. We also include the number
of double-bang eikonal events, which is significant for low
values of $M_D$.

The energy distributions of contained hadronic showers in
IceCube for both fluxes, $M_D=2$ TeV and $n=2\;(6)$ are also
shown in Fig.~\ref{fig6}.
There is a clear difference between the energy distribution 
of eikonal and BH or SM events: while these have a shape
similar to the cosmogenic flux, eikonal events are 
typically of much lower energies, specially for $n=2$.

\section{Summary and discussion}

Cosmogenic neutrinos interact with terrestrial nucleons
at center of mass energies
$\sqrt{2m_N E_\nu}\approx 100$ TeV, so they can be used as
probes of new TeV physics in neutrino telescopes. 
In particular, the possibility
of BH formation in models with extra dimensions has been
entertained by several groups.
These analyses are based on a geometric cross section
that assumes single BH production whenever the neutrino and the
parton interact at impact parameters smaller than $R_S$.
The problem with this estimate is that, despite the large
energy of cosmogenic neutrinos, the $\nu N$ cross section is
dominated by the small $x$ region and most of the BHs produced
in a neutrino telescope would be very
light, with masses just above $M_D$.
These light BHs would be very sensitive to effects like
graviton emission during the collapse or non-thermal effects 
in the evaporation, which add uncertainty to any estimate.

In this paper we have analyzed in detail a different 
type of signal. It is produced when the neutrino interacts
elastically with a parton at typical distances larger
than $R_S$ and transfers a small fraction $y$ of its
energy. The process is properly described by the eikonal
approximation. We have shown that the cutoff dependence 
of the eikonal cross section
is small (see Fig.~\ref{fig2}), and that non-linear corrections and 
graviton emission are negligible. The distinct 
experimental signature of these processes
would be a hadronic
shower of energy $yE_\nu\ll E_\nu$. A muon, a tau, or 
an electromagnetic 
shower would never be produced in the initial $\nu N$
interaction. 

If IceCube observes contained showers above 100 TeV, 
their energy distribution and the absence of 
charged leptons in the starting point of the shower
would then suffice to decide whether they
are due to SM or TeV-gravity interactions, as
we argue below.

Let us first suppose that the cosmogenic neutrino flux
(important at energies between $10^8$ and $10^{11}$ GeV)
is smaller than expected (see the {\it minimal}
flux in Fig.~\ref{fig3}). 
In this case there will be no SM contained
showers in IceCube. However, for sufficiently low values
of $M_D$ (the optimal value is 0.8 TeV, see 
$P^{eik}_{event}$ in Fig.~\ref{fig6})
the eikonal cross section grows
and even a very low cosmogenic flux could imply 
contained showers from gravitational interactions. These
showers would be typically 
less energetic than the initial neutrino that produce them 
(see distribution in Fig.~\ref{fig6}). 
One may then wonder if these
events could be due to a large flux of neutrinos in the 
intermediate energy range 
($10^5-10^8$ GeV). The observation of an initial muon
in a 24\% of the showers would be characteristic  
of SM interactions in this range of energies,
whereas the absence of muon events would be consistent
with {\it soft} gravitational interactions of cosmogenic 
neutrinos of much higher energy. 
In addition, the SM contained showers
would always come together with a large number of muon events
of similar energy produced outside the detector, which
would be absent for TeV-gravity contained showers.

Let us now suppose a flux of 
cosmogenic neutrinos within the expected limits 
(fluxes {\it higher} and {\it lower} in Fig.~\ref{fig3}). 
In this case IceCube 
will observe 
SM contained showers in the range $10^8$ to $10^{11}$ GeV.
If $M_D\approx 5$ TeV the number of eikonal processes will
be similar to the number of SM events. However, their energy
distribution will be different and both types of processes
can be separated. Again, the absence of charged leptons in
the initial interaction point would distinguish these events from
SM events of lower energies.
For the {\it higher} flux in 
Fig.~\ref{fig3} we expect more than one contained non-standard
shower per year in IceCube if 
$M_D\le 6.0\;(5.5)$ GeV for $n=2\;(6)$. 

We then conclude that for any intermediate-energy
and cosmogenic neutrino fluxes, 
an enhanced rate of neutral versus charged
current events of energies above 100 TeV could
be explained by TeV-gravity interactions. 

These interactions could also produce a very 
peculiar signal for relatively low values of $M_D$.
In a typical eikonal process the
neutrino loses a small fraction $y$ of its energy and
keeps going, so it can interact several times inside
the detector. This effect is specially important for
low values of $n$, where gravity dilutes slowly
with the distance (notice that for $n<2$  
it becomes a long-distance interaction 
and the total eikonal cross section is divergent).
If $n=2$ and $M_D\le 0.9$ TeV the 
average interaction length
of a $10^{10}$ GeV neutrino between two
eikonal interactions of $E>100$ TeV 
becomes smaller than the longitudinal dimension
of IceCube. Therefore, we would expect two (or more)
bangs of $10^5$ to $10^8$ GeV inside the detector. We think
this type of events could be easily distinguished from 
possible double-bang SM events, where the
first bang corresponds to a $\nu_\tau N$ charged current
interaction and the second one to a tau decay. First
of all, the typical energy of the SM double-bang event would be 
necessarily between $2.5\times10^{6}$ and $10^{7}$ GeV; for
lower energies the tau decays before 125 m (the separation
between strings at IceCube) and for larger energies the
second bang would be out of the detector.
Second, there should be a clear trace in the detector
as the tau
propagates between the two bangs, which is absent in
TeV-gravity events.

In summary, we think that elastic eikonalized 
interactions provide a clear (distinguishable from possible
SM events) and model-independent (insensitive to how the 
theory is completed in the UV) signal of TeV gravity. 
Being at impact parameter
distances larger than $R_S$, these 
interactions have a cross section that is larger than the geometric 
cross section to produce a BH. The eikonal event would be much
less energetic than a SM or a BH event, but neutrino 
telescopes are sensitive to showers of energy up to four orders
of magnitude below the average energy of cosmogenic neutrinos.
The values of the fundamental scale of gravity  
that IceCube could reach, around 5 TeV, are comparable to those 
to be explored at the LHC or the ILC \cite{riccardo}.

This work has been supported by MCYT (FPA2003-09298-C02-01) and
Junta de Andaluc\'\i a (FQM-101). J.I.I.~and D.M.~acknowledge 
financial support from the
European Community's Human Potential Programme HPRN-CT-2000-00149. 
We thank Eduardo Battaner, Tommaso Chiarusi, Francis Halzen, Marek
Kowalski, Paolo Lipari, Teresa Montaruli, Sergio Navas, Andreas 
Ringwald and Christian Spiering for useful discussions.


\begin{thebibliography}{99}

\bibitem{Anchordoqui:2002hs}
L.~Anchordoqui, T.~Paul, S.~Reucroft and J.~Swain,
Int.\ J.\ Mod.\ Phys.\ A {\bf 18} (2003) 2229.

\bibitem{semikoz}
D.~V.~Semikoz and G.~Sigl,
JCAP {\bf 0404} (2004) 003.

\bibitem{ADD}
N.~Arkani-Hamed, S.~Dimopoulos and G.~Dvali,
Phys.\ Lett.\ B {\bf 429} (1998) 263;
I.~Antoniadis, N.~Arkani-Hamed, S.~Dimopoulos and G.~Dvali,
Phys.\ Lett.\ B {\bf 436} (1998) 257.

\bibitem{BH0}
P.~C.~Argyres, S.~Dimopoulos and J.~March-Russell,
Phys.\ Lett.\ B {\bf 441} (1998) 96;
R.~Emparan, G.~T.~Horowitz and R.~C.~Myers,
Phys.\ Rev.\ Lett.\  {\bf 85} (2000) 499;
D.~M.~Eardley and S.~B.~Giddings,
Phys.\ Rev.\ D {\bf 66} (2002) 044011.
S.~Dimopoulos and G.~Landsberg,
Phys.\ Rev.\ Lett.\  {\bf 87} (2001) 161602;
S.~B.~Giddings and S.~Thomas,
Phys.\ Rev.\ D {\bf 65} (2002) 056010.

\bibitem{feng}
J.~L.~Feng and A.~D.~Shapere,
Phys.\ Rev.\ Lett.\  {\bf 88} (2002) 021303;
L.~A.~Anchordoqui, J.~L.~Feng, H.~Goldberg and A.~D.~Shapere,
Phys.\ Rev.\ D {\bf 65} (2002) 124027;
Phys.\ Rev.\ D {\bf 66} (2002) 103002;
Phys.\ Rev.\ D {\bf 68} (2003) 104025.

\bibitem{ringwald}
A.~Ringwald and H.~Tu,
Phys.\ Lett.\ B {\bf 525} (2002) 135;
M.~Kowalski, A.~Ringwald and H.~Tu,
Phys.\ Lett.\ B {\bf 529} (2002) 1;
S.~I.~Dutta, M.~H.~Reno and I.~Sarcevic,
Phys.\ Rev.\ D {\bf 66} (2002) 033002;
A.~Cafarella, C.~Coriano and T.~N.~Tomaras,
``Cosmic ray signals from mini black holes in models with extra dimensions:
An analytical / Monte Carlo study'', 
arXiv:hep-ph/0410358.



\bibitem{halzen}
J.~\'Alvarez-Mu\~niz, J.~L.~Feng, F.~Halzen, T.~Han and D.~Hooper,
Phys.\ Rev.\ D {\bf 65} (2002) 124015.

\bibitem{olinto}
E.~J.~Ahn, M.~Cavaglia and A.~V.~Olinto,
arXiv:hep-ph/0312249.

\bibitem{casadio}
R.~Casadio and B.~Harms,
Int.\ J.\ Mod.\ Phys.\ A {\bf 17} (2002) 4635;
D.~Stojkovic,
Phys.\ Rev.\ Lett.\  {\bf 94} (2005) 011603;
T.~G.~Rizzo,
``Collider production of TeV scale black holes and higher-curvature
gravity'', arXiv:hep-ph/0503163.
J.~L.~Hewett, B.~Lillie and T.~G.~Rizzo,
``Black holes in many dimensions at the LHC: Testing critical string
theory'', arXiv:hep-ph/0503178;
H.~Yoshino and V.~S.~Rychkov,
``Improved analysis of black hole formation in high-energy particle
collisions'', arXiv:hep-th/0503171.

\bibitem{Giddings:2004xy}
S.~B.~Giddings and V.~S.~Rychkov,
Phys.\ Rev.\ D {\bf 70} (2004) 104026.

\bibitem{emparan}
R.~Emparan, M.~Masip and R.~Rattazzi,
Phys.\ Rev.\ D {\bf 65} (2002) 064023;
M.~Masip,
arXiv:hep-ph/0210143.

\bibitem{us}
J.~I.~Illana, M.~Masip and D.~Meloni,
Phys.\ Rev.\ Lett.\  {\bf 93} (2004) 151102;
D.~Meloni,
Acta Phys.\ Polon.\ B {\bf 35} (2004) 2781.

\bibitem{eik}
G.~'t Hooft,
Phys.\ Lett.\ B {\bf 198} (1987) 61;
I.~J.~Muzinich and M.~Soldate,
Phys.\ Rev.\ D {\bf 37} (1988) 359;
D.~Amati, M.~Ciafaloni and G.~Veneziano,
Phys.\ Lett.\ B {\bf 197} (1987) 81;
D.~Kabat and M.~Ortiz,
Nucl.\ Phys.\ B {\bf 388} (1992) 570.

\bibitem{riccardo}
G.~F.~Giudice, R.~Rattazzi and J.~D.~Wells,
Nucl.\ Phys.\ B {\bf 630} (2002) 293.

\bibitem{icecube}
J.~Ahrens  [IceCube Collaboration],
arXiv:astro-ph/0305196;
see also http://icecube.wis.edu/.

\bibitem{colliders}
E.~A.~Mirabelli, M.~Perelstein and M.~E.~Peskin,
Phys.\ Rev.\ Lett.\  {\bf 82} (1999) 2236;
G.~F.~Giudice, R.~Rattazzi and J.~D.~Wells,
Nucl.\ Phys.\ B {\bf 544} (1999) 3;
for a review, see
F.~Feruglio,
arXiv:hep-ph/0401033.

\bibitem{SN}
S.~Cullen and M.~Perelstein,
Phys.\ Rev.\ Lett.\ {\bf 83} (1999) 268;
S.~Hannestad and G.~G.~Raffelt,
Phys.\ Rev.\ D {\bf 67} (2003) 125008
[Erratum-ibid.\ D {\bf 69} (2004) 029901].

\bibitem{Giudice:2004mg}
G.~F.~Giudice, T.~Plehn and A.~Strumia,
Nucl.\ Phys.\ B {\bf 706} (2005) 455.

\bibitem{peskin}
S.~Cullen, M.~Perelstein and M.~E.~Peskin,
Phys.\ Rev.\ D {\bf 62} (2000) 055012;
F.~Cornet, J.~I.~Illana and M.~Masip,
Phys.\ Rev.\ Lett.\  {\bf 86} (2001) 4235.

\bibitem{Antoniadis:1993jp}
I.~Antoniadis and K.~Benakli,
Phys.\ Lett.\ B {\bf 326} (1994) 69;
S.~A.~Abel, M.~Masip and J.~Santiago,
JHEP {\bf 0304} (2003) 057. 

\bibitem{Amati:1990xe}
D.~Amati, M.~Ciafaloni and G.~Veneziano,
Nucl.\ Phys.\ B {\bf 347} (1990) 550.

\bibitem{Lai:1999wy}
 H.~L.~Lai {\it et al.}  [CTEQ Collaboration],
Eur.\ Phys.\ J.\ C {\bf 12} (2000) 375.


\bibitem{Sreekumar:1997un}
P.~Sreekumar {\it et al.},
Astrophys.\ J.\  {\bf 494} (1998) 523.

\bibitem{Fodor:2003ph}
Z.~Fodor, S.~D.~Katz, A.~Ringwald and H.~Tu,
JCAP {\bf 0311} (2003) 015.

\bibitem{domokos}
G.~Domokos and S.~Kovesi-Domokos,
Phys.\ Rev.\ Lett.\  {\bf 82} (1999) 1366;
W.~S.~Burgett, G.~Domokos and S.~Kovesi-Domokos,
``Low scale string unification and the highest energy cosmic rays'',
arXiv:hep-ph/0209162;
Z.~Fodor, S.~D.~Katz, A.~Ringwald and H.~Tu,
``Strongly interacting neutrinos as the highest energy cosmic rays'',
arXiv:hep-ph/0310112.

\end{thebibliography}
\end{document}